# Effect of Key Match Events on Football Passmaps

Vid Stropnik[a,1], Vuk Đuranović[a], and Maruša Oražem[a]

[a]University of Ljubljana, Faculty of Computer and Information Science, Večna pot 113, SI-1000 Ljubljana, Slovenia

**A team of association football players may be envisioned as a directed network with player-nodes and weighted pass edges. Such a simplistic representation of an otherwise complex structure yields several benefits, but also permits the application of well-established network analysis approaches, to gain insights into the match flow. The authors of this work propose the application of these, including centrality measures, community detection methods and the analysis of induced subgraphs, to analyse potential effects certain match events have on the team. In addition to reporting on their findings, they emphasize their work's contribution of over 9000 novel football network representations to the Network Analysis community.**

Association football is the world's most popular sport. During gameplay, players attempt to create goal-scoring opportunities through various methods of individual control of the ball. While these can include dribbling, tackling and taking direct shots, the most common one is passing the ball to a teammate. Progressing the ball along the pitch with a series of passes helps the team retain control of the match, enabling them to realize set plays. It is worth noting that this method is not only common, but also effective, as exhibited by the correlation of per-game average pass frequency and overall tea, success. (1)

Ball movement and goal-scoring opportunities in a football match are complex phenomena to model. Pass characteristics, such as their length, speed or the involved players' positions are often used to provide further insights into their importance and quality. Logging such characteristics for hundreds of passes that happen in any given football match, however, renders the resulting data structure very complex and hard to interpret. To infer more general information about longer stints of play, simpler arrangements, such as networks with player nodes and directed pass links, should be considered.

There are several events in any given football match, that might incur a fundamental change in a team's tactics and approach towards the game. The coaching staff might recognize weak points in the opposition's strategy and relay them to the players during half-time. Suddenly conceding a goal in a must-win game might prompt a team to take more risks and try a different approach, while we can often observe leading teams resort to time-wasting tactics to ensure their victory.

In this work, we introduce novel ways of football match analysis using established network science approaches. By using methods of measuring change in team pass map centrality, by observing alterations of its underlying communities and by recognizing the (dis)appearance of distinct player graphlets as a result of a key event in a football match, we study common team behaviours that key match events might induce. We recognize these insights as useful in several areas. First, they serve as a form of establishing several ground truths about common trends of play, that will underpin further research. Secondly, more fine-grained difference of successful and unsuccessful responses to key events might be observed and be used in match preparations by domain experts. Finally, we recognize real-time applications of the proposed methods as useful in broadcasting, betting, commentary and analytics, to name only a few fields.

This report is structured as follows: In the following paragraphs, we provide a brief overview of the work, relating to inferring insights from sports data using network analysis. Here, we also introduce some key academic contributions to our work in the tree network science fields of special interest: measuring centrality, detecting communities and gathering semantic information from network graphlets. A subsection corresponds to each of these fields in the succeeding *Results* section. In them, we introduce the main revelations about the effects of key match events on passing trends. We later comment on the implications of these results and asses their appropriateness for further application, while finally introducing our detailed methodologies at the end of the paper.

## Related work

Evaluations of group performances based on their underlying network structures have been conducted for some time. Researchers were keen on understanding what kind of social, interaction or communication network structure boosts the productivity of a group in a working environment. Bavelas (2) was one of the pioneers in exploration of information transfer between groups. He discovered that the rate of information diffusion was significantly smaller in decentralized networks in comparison to centralized ones. It was later discovered by Shaw (3) that this gave an edge to decentralized groups over the centralized ones in solving complex tasks.

Team performance in collective sports has become a well researched area in the network analysis community in recent years. We recognize the conclusions by Grund (4) as a good starting point for the study of this field. In his work, he suggests that small networks (i.e. the eleven-node football team) are better explained with the use of the weighted edges between players. In such cases, weights are determined by the frequency of passes between two players. He also introduces a measure called network intensity, as a counterpart to network density in large and unweighted networks. This measure is computed as a function of weights for each link between players, nicely called the passing-rate of the team. As a result of his work, Grund confirmed that an increase in team passing rate leads to increased performance, while increases in centralization of play reduces team performance. Similarly to our work, network metrics were used to examine differences between halftimes in work by Clementine (5), albeit with a different form of pass-map representation. The authors of

---

([6](#)), again envisioned a football match as a weighted directed network, where they introduced a node measure called flow centrality - a function of pass accuracy and ratio of times a player was included in an action, leading to a shot on goal. Using it, they successfully graded individual player performances within a team. The measure is based on betweenness centrality, which has also been applied to football matches to examine correlations of physical demands in elite football players ([7](#)). While these works' conclusions are introduced as solid and objective, the importance of considering additional meta-data in the inference of flow using centrality measures was emphasized in the traffic domain in ([8](#)).

While several graphlet and motif-related methodology in our work is heavily inspired by the work of Pržulj ([9](#)), the concept of network motifs, initially shown by Milo et al. ([10](#)), has been successfully applied to sporting pass-maps by other authors as well. Notably, Penna and Touchette ([11](#)) analyzed the network motifs to analyze passing behaviour, while Bekkers and Shaunak ([12](#)) use meta-data equipped motifs (Possession Motifs and Expected Goal Motifs) to answer more in depth domain questions and graphically visualize team or player similarity by means of radar graph.

Due to advances in player-tracking technology, most recent research, such as those by Fernandez and Born ([13](#)), or by Chawla et al. ([14](#)), exploits the spatial-temporal data about observed passes. Due to time restrictions, posed by the gathered data, we don't consider these additional pieces of information in our analysis, but instead focus on the starting formation of the team in determining the positions of players. Researchers opting for the former approach of inferring information from additional metadata, however, should also consider work by Mattson and Takes ([15](#)) on pass trajectory extraction. Spatio-temporal methods are also successfully applied to football to accurately detect individual player behaviour and changes in team formation. ([16](#)).

### Results

**Centrality and Intensity.** Network centrality is one of the more common measures applied in Network science. In the context of sports passmaps, players with a high centrality value can be interpreted as those to whom team-mates often turn to; to organize play and carry the ball over congested areas of the pitch. Teams with a high variance in centrality scores are those who don't rely on individuals (usually industrial midfielders) to do this sort of heavy lifting, while others, where this deviation is low, tend to rely more on such star-players. Consider, for example, the two networks (Halftime snapshots of the same match), presented in Figure [1](#), where the difference between centrality variances is quite stark. To find out which key match events might induce such a change in centrality distributions, one might acknowledge the results introduced in Figure [2](#) - showing that player dismissal and the first goal in a match are two such events. The introduced results turned out to be surprisingly conclusive, with, of course, some expected outliers - the examination of which revealed unfavourable temporal splits (ie. goals scored very early, very late) as the cause of their deviation. Similar results were achieved when considering the flow centrality measure, while closeness centrality yielded null differences for all event types.

Another commonly used network science metric is network density. Usually applied to large, undirected networks, it is used to compute how many ties between network-actors exist, compared to how many ties between them are possible. Network intensity is an analogue of said approach for small, directed networks with a temporal component. Aptly named, it is an interesting measure of the team's pressing and dictation of the game's tempo, without other common domain metadata, such as the pitch-distance covered. In our results from Figure [3](#), we can notice changes of the size order of $\sim 15\%$ for certain key event splits. Along with the insights, provided in the Figure's caption, we can hypothesize that the post-dismissal network intensity increases, due to the fact that players have to cover more ground with the ball after losing a teammate.

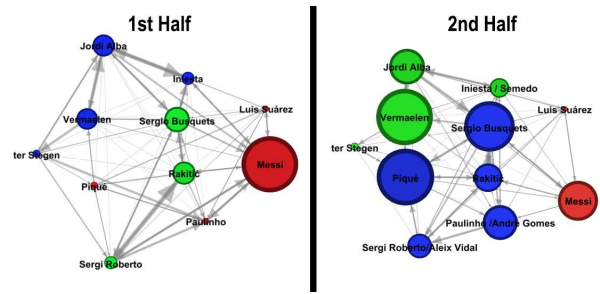

**Fig. 1.** Comparison of the two halftime networks snapshots from Barcelona's 2017/18 victory over Real Madrid. The left figure showcases the network before the halftime split, while the one after is shown on the right. Node size corresponds to their betweenness centrality, while colors denote communities, detected by the Leiden algorithm. Notice that the position of player centrality heavily shifts between the halftimes. Due to key events, happening in the second half, Barcelona were incentivized to attack less and to distribute their play across more players. In the second halftime, the standard deviation of betweenness centrality is increased by $\sim 50\%$.

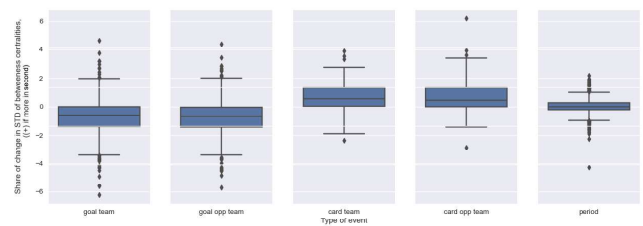

**Fig. 2.** The change in standard deviation of betweenness centralities in networks, observed before/after certain key match events. Positive values denote a larger deviation **after** the split, while negative ones mean more equal centrality measures before it. We can see that, no matter which team we observe, the first goal tends to induce more centralized, direct play in teams, which also conforms with the hypotheses, given in the caption of Figure [9](#). On the other hand, subplots 3 and 4 seem to indicate that the dismissal of a player in any team leads to the increase in variance.

**Community Detection.** One of our assumptions at the start of our work, was that the number of communities one might observe in these network snapshots, would change after a significant event. Throughout our research, we quickly noticed that, on such a small graph, it is hard to observe drastic and statistically conclusive changes in community numbers, as it is limited by the number of players that can be on the field in the same time. After testing all key events in order to check whether the number of communities changes, we concluded that no significant results were found. The most promising among all the considered events was player dismissal (the *red card* event), but even the exclusion of one player from the graph didn't bring about more communities in the network.

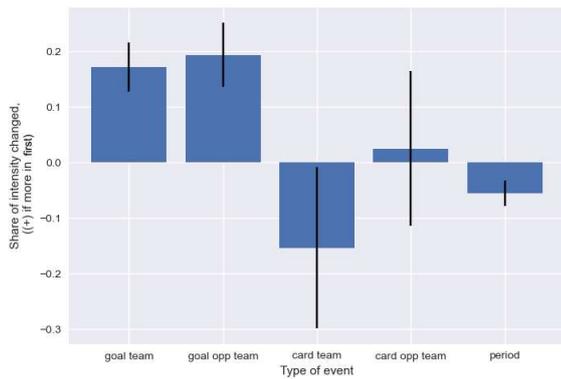

**Fig. 3.** The change in intensity, induced by different key match events. Notice that the match intensity seems to fall after the initial goal. We hypothesize that this might have something to do with the fact that the goal split is, from all of the observed ones, the one that tends to happen earliest in a match and might also be modelling player stamina.

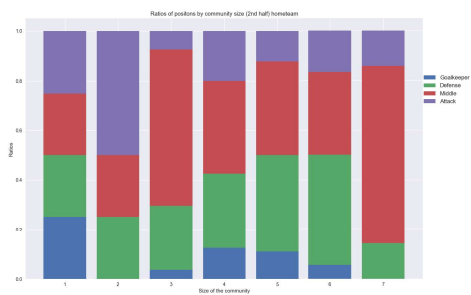

**Fig. 4.** Stacked barplot showing proportions of players by their position that appear in different communities. From this it can be observed that attackers tend to appear in communities of smaller sizes. Another observation to note is that larger communities will hardly exist without midfielders.

In contrast to that, we were able to find very interesting results regarding the structure of communities that occur in a game. It turned out that actual communities within a team's weighted graph differ very much from those pre-defined by position of a player on the field. As we initially intended to observe the proximity of communities to these playing lines across different event-splits, we measured their mutual information using hypothesized ground truths. After logging poor results for the supposed influence of important events on NMI (almost random behaviour was observed), we concluded that communities, discovered by detection algorithms, are different from traditional playing lines. This can also be observed in the communities of the example match in Figure 1. After acknowledging this information, we wanted to discover the source of this difference - What causes the detected communities to be different from those initially expected. While no conclusive results were achieved here, we did uncover the proportions of players by position, that appear in communities of different sizes, during our work. These insights, while *residual*, proved to be very interesting and are thus included in this report in Figure 4.

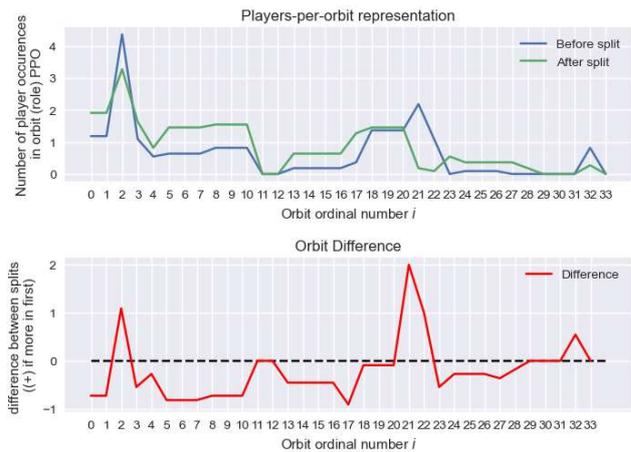

**Fig. 5.** The Orbit-occurence-per-player profile plots for Barcelona's 2017/18 victory over Real Madrid. The graph shows comparison between the two halves. We can see that the number of players occuring in the orbit number 21, as seen in figure 7 significantly decreased in the second half.

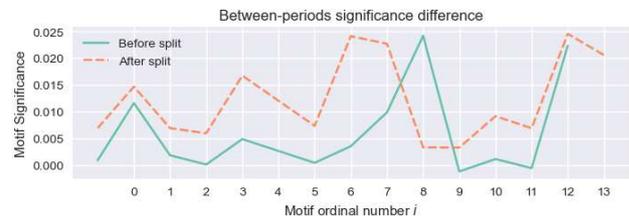

**Fig. 6.** The motif significance profile for Barcelona's 2017/18 victory over Real Madrid. From the graph, we can discern that the abundance of observed motifs *8* played a significant role in characterizing the first half, whereas there was a general increase in the significance of lower-order motifs in the second half.

**Network fragments.** Another popular technique for inferring global and local network information, is the analysis of its fragments. The concept of these induced subgraphs (motifs) and the roles (orbits) individual players represent within them, translates well to the consideration of a sports team's passing map.* Using these techniques, individual team performances may be profiled using orbit-occurences-per-player (OPP) or motif significance approaches. These, also visualized in Figures 5 and 6, offer a concise way of understanding differences between two passmaps - in these instances, from the same match. Combined with the knowledge of the chosen motif and orbit labelling conventions, here provided in Figure 7, such plots can be efficiently used to garner insights into the teams' dynamics.

Certain motifs characterize football ball movements better than others. In our general testing, as part of which we showed that no special change is induced by the half-time key event, we established the most notable ones to be the 0th, 7th, 8th and 13th. For all said motifs the median significance z-score exceeded the value of 3 standard deviations from the mean. Perhaps more interesting is the motif significance behaviour when observing the effect of a player-dismissal on a match. In expectation, the number of occurrences of the 9th motif will actually reduce significantly below the number of those, emerging from random 10 node configuration graphs. Along with this insight, Figure 8 also shows that the punished's opposition will actually attempt to exploit their numeric advantage using the playing patterns, previously established as most characteristic for football play passages.[†] This approach, and the analysis of orbits, prove to be somewhat interchangeable. Consider, for example, that the most significant nodes coincide with the orbits in which we can observe most key match event differences, such as those in Figure 9.

---

*Given that three to four player interlinks have been popularized both in pop culture and academia (17), with perhaps the most prominent representative being the infamous Triangle offence in American basketball.

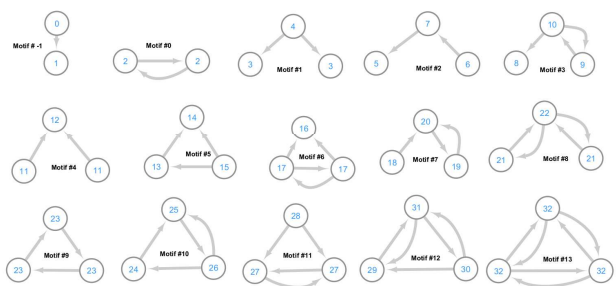

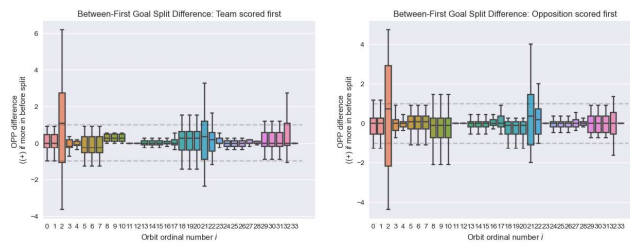

**Fig. 7.** The generated directed motif atlas, denoting all possible induced subgraph types of sizes 2 and 3. The nodes are labelled with the 33 possible orbital roles.

**Fig. 9.** The distribution of values of *before - after* Orbit occurences per player differences, observed for the *First goal* key event split. With positive values denoting a higher amount of OOP before the initial goal, we can notice that the OOP in orbits #2 and #21 tends to go down after the initial goal for both teams. Since these two orbits can be interpreted as very organized and defensive, a possible explanation for this is *the breaking of the deadlock* with the first goal: the initial goal prompts players to abandon more conservative plans and play more expressively, no matter which team took the lead.

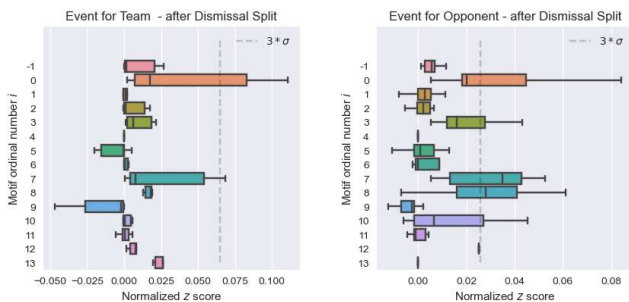

**Fig. 8.** The motif significances in graphs succeeding player dismissal for the observed team (left) or their opponent (right). Notice that certain less common passing trends, such as the one observed in motif #9 in Figure 7 tend to become so rare that they actually appear more commonly in configuration graph models than in the networks of punished teams. This probably due to the fact that teams with less players tend to express themselves mostly in the more characteristic motifs, such as the more significant #0 and #7. Meanwhile, these characteristic models tend to express themselves very clearly for the opposition team, who is expected to be in control in this network snapshot.

## Discussion

Several key interpretations of our results are provided in the captions of the figures of this report. In them, we show, on multiple occasions, that network analysis approaches are not only very suitable for discovering interesting truths about football events, but are also tend to be moderately conclusive in non-trivial insights. Consider, for example the findings about intensity and centrality deviations from Figures 2 and 3, achieved without abundant amounts of metadata,[‡] generally available in this domain. Consequently, a big contribution of our work is the showcase of the applicability of very simplistic methods for the inference of certain match characteristics. Secondly, we see our work in garnering passmap insights using its induced fragments and their corresponding orbits as directly applicable for use in sports analytics. The profile plots, such as those, shown in Figures 5 and 6, are not only informative on their own, but conform well to several standard similarity and agreement measures and are thusly well suited for inferring similar play styles or understanding refined semantics of their differences. When observing Figure 7, we can easily apply our results to the domain, to understand the important motifs as those denoting buildup play, with motif #8 being a clear representation of a single-pivot playing shape (with the pivot being the player in orbit #22). These observations are important, because they open the possibility of dimensionally reducing the profile plots, rendering potential analysis even easier. Our overall goal for this project was to understand how key events might affect football pass-maps. While nothing conclusive can be said, due to our three events naturally being based on overlapping data, many interesting results were found. A perhaps shocking revelation, and one that should be studied further, is the emergence of similar behaviours in both networks, no matter if the key event was in favour or against any given team - whereas one might expect an inverse to be the more natural response in such a situation. The final, and perhaps most significant contribution of our work, is the repository (18), in which all of our work is shared. Along with the source code for the replication of our analyses and additional visualizations, one can find the parsed networks for almost 800 matches available for download. Considering two teams per match and 6 networks per team, our work has contributed over 9000 unique networks in pajek format with metadata documentation to the open source network analysis community.

## Methods

**All of the described results were achieved by analysing directed networks with player-nodes and weighted pass edges, with weights corresponding to pass frequency.**

**Data.** All of the examined networks were parsed from alternative data structes by the authors of this work. The original event files were provided by Statsbomb (19). Unless noted otherwise, the described results were achieved by aggregating the findings of analyses, conducted on 789 parsed matches. For each successful pass, a directed edge was generated between the players, logged in the corresponding pass attempt and reception events by the data provider. For the three considered key events - the half time break, the first goal and the first dismissal in any given match, 6 directed *pajek* files were created for each involved team (3 before and 3 after the key event), where games without certain events contain empty *after-event* files. Additionally, the graphs' flow centrality variants were created and saved according to the procedure, described in the following subsection. For the calculation of network intensity, the description of which also follows, the team posession time was calculated as the sum of timestamp differences in each uninterrupted posession stint.

---

[†] Due to dismissals usually happening late in the analyzed graphs, not enough data was gathered to infer the correct significance of the 13th component, while it additionally being really rare in configuration model networks.

[‡] such as players' spatial location, useful in determining passing lanes, or pedometers and radars, which are used to compute measures, similar to intensity.

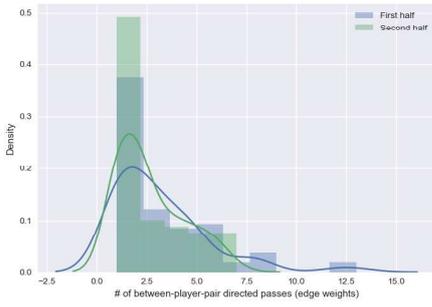

**Fig. 10.** The distribution of graph edge weights in Bayern Münich's 2012 Champions league final loss versus Chelsea shows that edge weights roughly follow a Poisson distribution. As the same trends emerge in other examined games as well, we note that pruning the graphs at just above the median $\lfloor \lambda + \frac{1}{3} - \frac{0.02}{\lambda} \rfloor \approx \lambda$ will result in considering only the non-trivial passes, carrying information about frequent passes of play.

**Centrality and intensity.** Three centrality measures were calculated on each network. These include the well known closeness and betweeness, as well as a modified version of flow centrality, initially introduced in (6). All tree centralitiy distributions were calculated on all 789 networks. Then, differences in standard deviations on each key event split were calculated. While the first two measures were calculated on regular network snapshots, the modified flow centrality utilized networks, specially curated for this task. These included two additional nodes, representing shots made on and off goal. Our modified flow centrality score is, mainly due to scarce methodological descriptions in the original work, simply analogous to the calculation of betweenness centrality on these special networks. In all inference about individual player centrality scores and their deviations, the two custom nodes were always skipped.

Additionally, for each event and for each team, the intensity measure was calculated. As originally introduced in (4), it is defined as

$$I(team) = \frac{1}{T} \sum_i \sum_j w_{ij},$$

where $T$ is the cumulative time the observed *team* posessed the ball, and $w_{ij}$ is the weight between node $i$ and $j$.

**Community Detection.** To find sensible semantic groupings in our weighted graphs, we mainly used the Leiden community detection algorithm. Prior to performing the search, we pruned the networks, so that edges, weighing less or equal to the median edge weight in the examined graph, were removed. The estimation of an adequate pruning point stems from the insights, described in the caption of Figure 10. In order to check for any valuable information we may have missed with our algorithm selection, we ran additional experiments using the clique percolation method.

All methods, in which player formation metadata was used, required us us to manually input correct formations (line-ups) of teams in the observed matches. An example of such was the task of determining the extent to which the communities, detected in certain periods of a game, differ from standard playing lines. Since this process of manual input is a very time-consuming task, we restricted ourselves to the analysis of 20 matches. For both teams in these matches, we programatically described the ground truth formations as a set of nested lists, similar to the formats returned by some community detection algorithms.

Similarity between the detected communities and communities pre-defined by position was measured using NMI (normalized mutual information). The ground truhts of player positions were inferred using *google.com*'s formation search feature. Our problems with this approach, described in the results section, mainly stemmed from our initial hypothesis and not the used methodology or tools. On the other hand, we also encountered a problem when using the Leiden detection algorithm. In some cases, not all nodes were classified into communities. The yield of the detection method was only comprised of those nodes, that appear in a community at least with some other node. In such cases, nodes that were not assigned to any community, were classified as singletons.

**Network fragments.** In all experiments concerning network fragments, the networks were pruned before examination, as already described in the prior subsection. Due to the time complexity of the tasks involving the generation of random subgraphs, we limited ourselves to the analysis of 32 matches for those approaches.

To examine the orbit-representation in the observed graphs, an atlas of all possible directed graphlets of node-count 3 or less was created. Here, the empty motif and singelton were omitted, resulting in the collection from Figure 7. From there, each unique edge subset from the analysed graph was extracted, and a hash table of benchmark orbital roles created. For each subgraph, inferred from the edge subsets, the orbital roles for each contributing node were logged. The resulting table of this procedure denoted the number of occurrences of each player in each of the 33 considered orbital roles. Each network's orbital profile was then calculated as the cumulative number of player-occurrences in each orbital role, divided by 11 - the number of players on the pitch. By comparing the differences between these profiles before and after key events, the final insights of the distrbution of OOP differences, was achieved.

Similarly, the induced motifs were also initially counted by detecting isomorphisms between fixed-size sub-graphs and graphlets from the atlas. In an approach described in (9), the significance of each motif is calculated by comparing the number of its incidences to that expected in a directed configuration model, corresponding to the considered graph's in and out degrees. In our approach, 100 different configuration models are created to compute these expected values. From there, the motif significance profile for motif $i$ is calculated as shown in Equation 1.

$$SP_i = \frac{Z_i}{\sqrt{\sum_i Z_i^2}} = \frac{\frac{n_i - E[\tilde{n}_i]}{\tilde{\sigma}_i}}{\sqrt{\sum_i (\frac{n_i - E[\tilde{n}_i]}{\tilde{\sigma}_i})^2}}, \qquad [1]$$

where $n$ is the number of motif occurrences, $\tilde{\sigma}$ is the standard error of all empiric configuration graph observations of $\tilde{n}$, the number of motif occurrences in the configuration model.